\documentstyle[12pt,amssymb,epsfig]{article}
\author {Ernesto S. Loscar, Rodolfo A. Borzi and Ezequiel V. Albano$^{a}$\\
$^a${\it Instituto de Investigaciones Fisicoqu\'{\i}micas
Te\'{o}ricas y Aplicadas}\\{\it (INIFTA), UNLP, CONICET,
Suc.4, CC16,}\\{\it
1900 La Plata, Argentina}}

\title{Interplay between thermal percolation and jamming
upon dimer adsorption on binary alloys.}

\begin{document}
\maketitle

\begin{abstract}

Using Monte Carlo simulations we study jamming and percolation processes upon the random sequential adsorption of dimers on binary alloys with different degrees of structural order. The substrates we use are equimolar mixtures obtained utilizing the isomorphism
between an alloy and the Ising model with conserved order
parameter. We anneal the substrates at temperature $T$ until they 
reach thermal equilibrium, and then we quench them to freeze
the state of order of the alloy at that temperature for the posterior adsorption trials. The annealing
temperature is then a continuous parameter that characterizes the
adsorbing surfaces, shaping the deposition process. As the
quenched alloys undergo an order-disorder phase transition at the
Onsager critical temperature ($T_{c}$), the jamming and
percolating properties of the set of deposited dimers 
are subjected to non-trivial changes, which we summarize in a 
density-temperature phase diagram. We find that for $T < T^* = 1.22 T_{c}$
 the occurrence of jamming prevents the onset of
percolating clusters, while percolation is possible for  $T > T^{*}$.
Particular attention is focused close to $T^{*}$, where
the interplay between jamming and percolation restricts
fluctuations, forcing exponents seemingly different from the
standard percolation universality class. By analogy with a thermal
transition, we study the onset of percolation using the {\it
temperature} (in this case, the substrate annealing temperature)
as a control parameter. By proposing thermal scaling Ansatzes we
analyze the behavior of the percolation threshold and its
thermally induced fluctuations. Also, the fractal dimension of the
percolating cluster is determined. Based on these measurements and
the excellent data collapsing, we conclude that the universality
class of standard percolation is preserved for all temperatures.
\end{abstract}

\newpage

{\bf I. Introduction.}
\\
\\
When adsorbed particles are bound to a solid surface the magnitude of the interactions
relative to the thermal energy may lead either to reversible or irreversible attachment to the surface.
If the particle-substrate interaction is weak enough, the particle can explore the whole surface through surface diffusion and desorption.
These mechanisms may eventually enable the system to reach equilibrium. The phenomenon could then be investigated using the methods of
equilibrium statistical mechanics. On the other hand, in the case of irreversible attachment
due to strong particle-substrate interactions, the particles once fixed on a surface neither
desorb from it nor diffuse on it. This irreversible adsorption or \emph{deposition}
of particles onto a solid substrate is a far-from equilibrium phenomenon of wide
interest in physics, chemistry, biology, and in other branches of science and technology.
Some examples embracing deposition include adhesion of colloidal particles and proteins,
separation of viruses or bacteria,  adsorption of gas molecules \cite{reviews,evans}, etc.
In addition to the usual case where homogeneous surfaces are used, the deposition can also be
performed on substrates where localized adsorption takes place on particular sites of non-homogeneous  substrates \cite{reviews,polaco}.
The case of deposition of gas particles on a substrate is of great practical importance as a
first step for chemical reactions in heterogeneous
catalysis \cite{albanol}. There, the substrate properties are usually improved by alloying,
and a large dependence of the catalytic properties on the composition and the configuration
of the surface can be observed \cite{libros}.

The random sequential adsorption model \cite{evans,priv} (RSA) provides an excellent description of the process of deposition, assuming the successive adsorption of particles within a lattice gas framework.
Within this model, objects of finite size are randomly adsorbed on an initially empty $d$-dimensional
substrate with the restriction that they cannot overlap with previously deposited objects.
The state of a site then changes irreversibly from empty to occupied. Under these conditions
the system evolves with a dynamics that becomes essentially dominated by geometrical
exclusion effects between particles. During particle deposition one can define different clusters looking for the sets of neighboring
occupied sites. A particular cluster is said to be
\textit{percolating} if it reaches two opposite edges of the lattice (\textit {e.g.} top and bottom).
The lowest coverage at which there is a percolating cluster on the infinite lattice is called the
\textit{percolation threshold} $\theta_P$. Since no desorption is allowed, the deposition process necessarily ends due to blocking, when no more particles
fit in the volume; in this context we say that \textit{jamming} occurs. The fraction of total space  covered at time $t$ by deposited particles, $\theta(t)$, reaches then a maximum value
$\theta(t\rightarrow\infty)=\theta_{J}$ called the \textit{jamming coverage}.

The saturation or jamming of a volume is an old issue that is still important today,
linked to a wide variety of problems as relevant as car parking, occupied volume
fractions on glasses and liquids, or packing of commercial granular goods,
for which is still the focus of great attention \cite{evans,Don_04,Mak_02}.
On the other hand, percolation is one of the most fundamental and widely
studied topics in statistical physics. The concept is applied to many problems of completely different types of fields ranging from natural sciences to sociological phenomena. The infection of trees in an orchard \cite{Stuffy}, magnetism on diluted alloys \cite{Ber_04}, conductivity on complex oxides \cite{Mot_03}, and the spread of forest fires \cite{perc2} are some popularly mentioned examples. For reviews on percolation see \textit{e.g.} \cite{Stuffy,Feders,Havlin} . As in the case of \textit{thermal} transitions, the percolating transition presents a non-trivial critical behavior, but due to purely geometrical causes. It shows scale-invariant behavior characterized by critical exponents with scaling relations between them. Furthermore, these exponents
exhibit universality and do not depend on microscopic details such as the inclusion of next-nearest-neighbor connectivity, or the nature of the lattice \cite{Stuffy,Feders}.

The connection between both phenomena---jamming and percolation---has been attracting considerable attention \cite{lero,choi,van,kon,kon1,fede,lba1,lba2} and it has been shown that they share some similarities \cite{kon,kon1,lba1,van}.
In some models the deposited objects cannot percolate because jamming occurs
before, blocking the system \cite{naka,kon1}. It has also been shown
that a continuous control parameter can be tuned to enforce the jammed system to go from a percolating region to a non-percolating one \cite{evans2,choi,fede,giorgio,polaco1}.

In the present paper we study the RSA of dimers on substrates composed of two-dimensional binary alloys \cite{lba1,lba2} using Monte Carlo
simulations. We first obtain the jamming coverages and the percolation thresholds by proper extrapolation to the thermodynamic limit. The temperature at which we have prepared the alloys constitutes a continuous variable that allows us to vary both the connectivity and the maximum coverage of the substrates. Using this temperature as a control parameter we can finely tune the jamming threshold in order to force its interference with the percolating transition. We can also generalize standard scaling concepts to a new scenario in which a thermal
parameter intervenes in a geometrical transition \cite{cecilia}. All these proposals are tested by means of computer simulations.

The manuscript is organized as follows: in Section II the models for
the substrate and adsorption process are defined and the simulation method is described. In Section III results are presented and discussed, while in Section IV the concept of thermal percolation is proposed, tested and discussed. Finally our conclusions are summarized in Section V.
%
\\
\\
{\bf II. Model and simulation method.}
\\
\\
We study the random sequential adsorption of dimers---i.e. two identical units---on
inhomogeneous substrates. The surface used for each deposition is a two-dimensional alloy annealed at temperature $T$ and then suddenly quenched to freeze the high-temperature configuration. We obtained different microstates of the alloy by means of Monte Carlo simulations on a square lattice of side $L$, using periodic boundary conditions and Kawasaki dynamics. We took advantage of the well known isomorphism
between the Ising model \cite{Binder} and a binary alloy, namely spin-up $\rightarrow A$-species and spin-down $\rightarrow B$-species, keeping the same density of particles
$\rho_A=\rho_B=1/2$. The ``annealing" temperature at which the substrate was generated
is measured in units of the interaction constant ($J$),
setting the Boltzmann constant to unity. We assumed attractive interactions
between species of the same type (\textit{i.e.} $J>0$, corresponding to
the ferromagnetic Ising model). It is well known that this system
undergoes an order-disorder transition at $T_C\simeq2.269$ in two dimensions \cite{Binder}. Since achieving equilibrium is particularly difficult at low temperatures, we choose as an initial condition the microstate that minimizes the energy (two rectangular domains of $A$ and $B$-atoms, respectively) to save computational time. We disregarded a high number of configurations correlated with the initial to ensure that equilibrium was attained
for each value of $L$ and $T$. For big lattices this demanded discarding more than $10^5$ Monte Carlo steps near $T_C$ and at the lowest temperatures. In order to perform the RSA experiments we generated and stored between 100 and 500 well equilibrated configurations of the alloy
(depending on the substrate size) for each annealing temperature, quenching in this way the state of order the substrate had at this $T$.
Subsequently, we study the
irreversible deposition process on top of the different substrates at zero temperature ---$i.e.$
neglecting the diffusion of the adsorbed dimers. The only relevant temperature we will be refereing to throughout this work is then the one at which the adsorbing surfaces have been prepared.
The RSA rule we assumed is the following: dimer adsorption on the alloy is only possible
on nearest-neighbor sites with atoms of \textit{different type}
($AB$ pairs), and it is rejected otherwise.

Simulations were performed by using samples of side $16\le L\le 512$, where distances are measured in lattice units. For additional details on the simulation method see reference \cite{lba1}.

Throughout a RSA process, the probabilities to find a percolating (jammed) cluster, on a finite sample of side $L$, can be fitted by the error function \cite{van,stuffyb}

\begin{equation}
P^{x}(\theta) = \frac{1}{\sqrt{2\pi}\sigma_{x}(L)} \int_{-\infty}^{\theta} 
exp\Big{[}-\frac{1}{2}
{\Big{(}\frac{\tau - \theta_{x}(L)}{\sigma_{x}(L)}\Big{)}}^2
\Big{]}d\tau,
\label{error}
\end{equation}
\\
\noindent where $\theta$ is the density of adsorbed dimers on the binary alloy, $\theta_{x}(L)$ is its mean value and  $\sigma_{x}(L)$ is the fluctuation of that density,
and $x=J,P$ refers to the jammed or the percolating state, respectively.

The binary alloy is a non-homogeneous substrate with a characteristic
structure determined by the thermal noise during the annealing period, while the adsorption of dimers is another random process. So, we have to deal with two correlated stochastic
processes and the measurement of relevant physical quantities requires
a careful treatment. In fact, if one has series of $n$ independent
samples of the substrate ($i=1,..,n$), it is possible to obtain
representative values of the coverages and their fluctuations by computing the following averages on samples obtained at $T$

\begin{equation}
\theta_{x}= \sum_{i=1}^{n}\frac{\theta_{x}^{i}}{n} ,
\label{thetax}
\end{equation}

\noindent and

\begin{equation}
\sigma_{x}= \sum_{i=1}^{n}\frac{\sigma_{x}^{i}}{n} .
\label{sigmax}
\end{equation}

It should be remarked that the sets ${\theta_{x}^{i}}$ and ${\sigma_{x}^{i}}$,
are obtained by making several adsorption trials ($\sim10^3$) using a
\textit{single} substrate and by calculating the average of the density and
its root mean square (RMS) over these trials. It has already
been shown that the measurement of the observables
defined by means of Equations (\ref{thetax}) and (\ref{sigmax})
captures the physical behavior of the adsorption
process \cite{lba1} (in the present case percolation and/or jamming).
In this way, for each substrate we perform several trials searching
first for percolation and after that for the jammed state. On other hand,
the fluctuations of the quantity $\theta_{x}^{i}$ taken over
$i$-indices, reflects the physical behavior of the substrate. This
kind of study has already been performed \cite{lba1}, so it will
not be repeated here.

\begin{figure}
\centerline{
\includegraphics[width=7cm,height=5cm,angle=0]{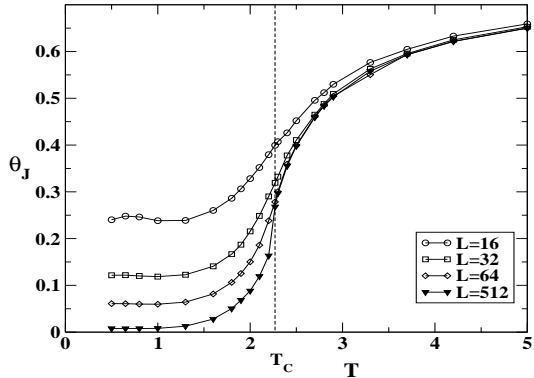}}
\caption{Plots of the jamming coverage ($\theta_{J}$) {\textit versus}
the temperature $T$ at which the substrates were prepared and then quenched from. As throughout the manuscript, we performed the depositions considering no thermal agitation at all. The vertical dashed line shows the location of the ordering temperature of the underlying alloy. The coverage, obtained for the RSA of dimers on lattices of different size $L$, closely follows the behavior of the energy of the underlying Ising model, with a sharp slope at $T_c$. Finite size effects are more evident for $T < T_c$; in this range of temperatures the deposition takes place mainly on domain walls.}
\label{Fig1}
\end{figure}

Another method for obtaining both $\theta_{x}^{i}$ and $\sigma _{x}^{i}$
is to directly fit an error function (Equation (\ref{error})) to the
data. Although this procedure is more demanding from the
computational point of view it will allow
us to take into account the interplay between percolation and jamming
processes explicitly in the evaluation of the relevant quantities
(for example critical exponents).
\\
\\
{\bf III. Results and discussion.}
\\
\\
{\bf Jamming}
\\
\\
Figure \ref{Fig1} shows plots
of the jamming coverages $\theta_J$ {\textit versus} $T$ obtained for samples of different size. These curves depend markedly on the temperature.
Indeed, given the deposition rule that we imposed for the dimers the behavior of
the saturation coverage qualitatively follows that of the energy
of the binary alloy used as a substrate, with a steep slope at $T_c$ (recall that the alloy energy is proportional to the number of broken bonds or $AB$-pairs).
In addition to this strong variation with temperature (from near zero to above 65$\%$ of the entire lattice) the jamming coverage presents strong finite-size effects for $T<T_C$.
It is well known \cite{evans} that the fluctuations of the
jamming coverage ($\sigma_{x}(L)$ in Equation (\ref{error}) with the subindex $x \equiv J$)  scale with the lattice size according to

\begin{equation}
\sigma_{J} \propto L^{-1/\nu_{J}} ,
\label{sigmaLJ}
\end{equation}

\noindent where $\nu_{J}$ is the jamming exponent. A relationship similar
to Equation (\ref{sigmaLJ}) also holds for the fluctuations of the percolation threshold
\cite{Stuffy}. In recent papers \cite{lba1,lba2} we have proved rigorously and tested by means of Monte Carlo simulations that Equation (\ref{sigmaLJ}) holds for a wide variety of RSA processes with a jamming exponent given by

\begin{equation}
\nu_{J} = \frac{2}{2D - d_{f}} .
\label{nuJ}
\end{equation}

\noindent Here, $D$ is the dimensionality of the space and
$d_{f}$ is the fractal dimension of the subset of active sites,
\textit{i.e.} sites that can allocate dimers.

\begin{figure}
\centerline{
\includegraphics[width=7cm,height=6cm,angle=0]{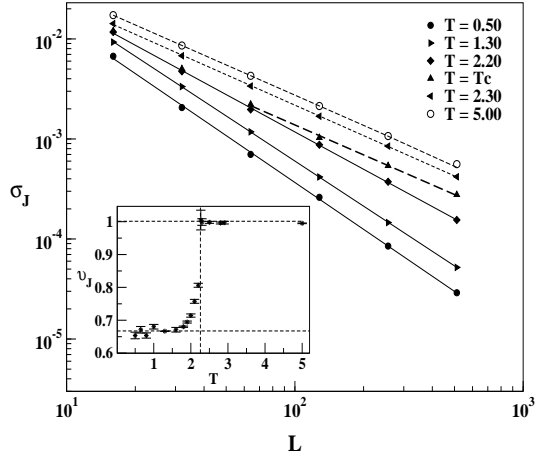}}
\caption{Log-log plots of the variance of the jamming coverage ($\sigma_{J}$) {\textit versus.} $L$, obtained after deposition at different annealing temperatures $T$ of the substrate. The inset shows the temperature dependence of the exponent $\nu_J$ obtained by fitting the curves to Equation (\ref{sigmaLJ}). The redistribution of atoms in the alloy at $T_c\approx2.269$ leads to a rounded step in $\nu_P$. This exponent, in turn, allows us to evaluate an effective dimensionality for the set of deposited dimers.}
\label{Fig2}
\end{figure}

In Figure \ref{Fig2} the log-log plots of $\sigma_{J}$ \textit{versus} $L$ show that Equation (\ref{sigmaLJ}) actually holds for the whole inspected range of annealing temperatures $T$ and lattice sizes. By fitting the data obtained at temperatures well below $T_{c}$ ($T < 2.0$), we have
determined $\nu_{J} \simeq 2/3$ (see the inset of Figure 2).
This result is in agreement with the idea that at very low temperatures the RSA process is essentially restricted to
a one-dimensional interface between well-conformed domains of different atoms. In this case one has $d_{f} = 1$ and Equation (\ref{nuJ}) predicts
$\nu_{J} = 2/3$ in $D = 2$ dimensions.

Despite this good agreement, we will argue that the true dimensionality of the \textit{whole} set of sites where dimers have adsorbed is not equal to 1.
When the temperature rises from zero it becomes increasingly probable that unlike species in the alloy start to diffuse from the domain wall into the bulk. They conform islands of one or more atoms surrounded by a sea of atoms of the other type, with their shores providing $AB$ pairs suitable for the adsorption of dimers. These islands should be present for \textit {any} non-zero temperature, increasing in number as $L^2$.
It is worth noting that these simple geometries are jammed by a fixed number of dimers: even though they add to the coverage they give no contribution to the fluctuation $\sigma_{J}$.
In the low temperature range and for the sizes analyzed,
we cannot see the large concentration of islands needed to have interference among them---or with the domain walls---and so the leading contribution to $\sigma_{J}$ is \textit{effectively} one dimensional. In fact, in Figure \ref{Fig2} we can see that above this temperature range and close enough to $T_{c}$ the interference effect starts
generating a rather smooth increase of $\nu_{J}$ when we approach the critical point from below, instead of a sharp step (see the inset of Figure \ref{Fig2}).
In this way, $d_f$ varies smoothly over \textit{effective values} between one and two.

Finally, if we look now for $T \geq T_{c}$, the long domain walls between $A$ and $B$ phases have disappeared (the disordering trend of the temperature has overcome the ordering tendency of the interactions among atoms) and one has that adsorption sites are almost  homogeneously distributed on the sample with $D = 2$ and $d_{f} = 2$, so that Equation (\ref{nuJ}) yields $\nu_{J} = 1$ in excellent agreement with the numerical results (see the inset of Figure \ref{Fig2}).

In view of the previous analysis and in order to extrapolate the jamming
coverage to the thermodynamic limit ($L\rightarrow\infty$), we
propose an Ansatz based on the assumption that
$\theta_{J}(L)$ has two leading contributions: i) the first one
corresponds to dimers adsorbed in the two-dimensional bulk
($\theta_{J}^{B}$), which is independent of $L$; and ii) the
second term ($\theta_{J}^{Int}$) arises due to the adsorption of dimers along the
interfaces between domains of different species, with \textit{effective dimension} $d_{f}$ which depends on $L$ as a power law $\theta_{J}^{Int} \propto L^{(d_{f} - D)}$. Then

\begin{equation}
\theta_{J}(L) = \theta_{J}^{B} + A L^{-(D - d_{f})},
\label{estrapoteta}
\end{equation}
\noindent where $A$ is a constant. Note that equation (\ref{estrapoteta}) resembles the scaling law generally used for the percolation coverage (see Equation (\ref{estrapolar5}) below).
\begin{figure}
\centerline{
\includegraphics[width=7cm,height=5cm,angle=0]{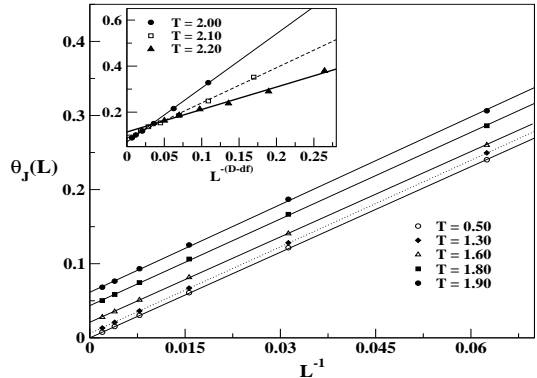}}
\caption{ Plots for the jamming coverage $\theta_{J}(L)$ {\textit versus} $L^{-(D-d_f)}$
for substrates at different temperatures $T$. We took the values for the dimension $d_f$ from the analysis of the fluctuations in the coverage $\sigma_J$ (Figure 2). The main figure condenses some results at low temperature, at which $d_f = 1$, while in the inset we restrict the temperature range to $2.00\leq T < T_c$, where $d_f$ departs from this value. The non-zero interception with the vertical axis indicates that there are dimers adding to the jamming coverage but not to its fluctuations. This implies that $d_f$ is only an effective dimensionality for the set of deposited dimers.}
\label{Fig3}
\end{figure}
Figure \ref{Fig3} and its inset show plots of $\theta_{J}(L)$ {\textit versus} $ L^{-(D-d_f)}$ for various temperatures below $T_c$ ($T\leq 2.20$). The value of $d_{f}$ that we used was obtained by inserting the effective exponents $\nu_{J}$, shown in the inset of Figure \ref{Fig2}, into Equation (\ref{nuJ}). The quality of the linear fits indicates that Equation (\ref{estrapoteta}) holds over the whole range considered (even for temperatures near but lower than $T_c$, as shown in the inset) according to what was discussed above, with effective dimension $1 \leq d_{f} < 2$. The second term of Equation (\ref{estrapoteta}) vanishes for $L \rightarrow \infty$ and the interception with the vertical axis provides us with an estimation of the jamming coverage in the thermodynamic limit, namely $\theta_{J}(L \rightarrow \infty) = \theta_{J}^{B}$ for $T < T_{c}$. We stress again that this limit would be 0 if $d_f$ were the true dimension of the set of active sites.

\begin{figure}
\centerline{
\includegraphics[width=8cm,height=6cm,angle=0]{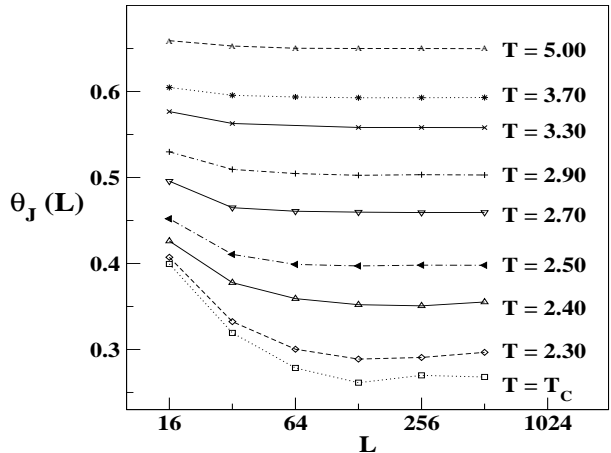}}
\caption{Linear-logarithmic plots showing finite size effects for the jamming coverage $\theta_{J}$ as a function of $L$ in a range of temperatures ($T$)
above the ordering temperature of the substrate. $\theta_{J}$ saturates for relatively small sizes, due to the two-dimensional distribution of the adsorbed dimers at high temperatures. Also, finite-size effects become less important for $T$ much higher than the ordering temperature of the alloy ($T_c$).}
\label{Fig4}
\end{figure}

Finally, for $T \geq T_{c}$ the jamming coverage only depends on
the system size for very small lattices reaching a stationary
value even for modest lattice sizes, as shown in Figure
\ref{Fig4}. 
This fact reflects the negligible operation of
lattice-size effects on the density of $AB$ pairs in the bulk of
the binary alloy above criticality. This finding could be
anticipated after inspection of Figure \ref{Fig1} and has also
been considered in the formulation of the Ansatz given by Equation
(\ref{estrapoteta}) since for $T \geq T_{c}$ one has $D=d_{f} = 2$.
\\
\\
{\bf Percolation}
\\
\\
Before analyzing percolation in depth, it is worth mentioning that the maximum density of adsorbed dimers obtained in this inhomogeneous RSA process is very low, particularly at temperatures below criticality (see Figure 1). While we only have $\theta_J\simeq0.5$ for $T \simeq 2.80$, this density
further decreases at lower temperatures. If we consider random percolation of dimers in the homogeneous case, the percolation threshold for the incipient percolating cluster is
close to  $\theta_{P} \simeq 0.56$ \cite{van}. Naturally, we may note that not only the density but also the geometry is an important factor, and that the elongated shape of domain walls present in the alloy may serve to nucleate percolating clusters at low temperatures, in spite of the low coverage.
Still, in our simulations we have found that percolation of dimers is not possible for $T < 2.80$. Of course for $T = 0$ with $\rho_{A} = \rho_{B} = 1/2$ one has that the ground state of the
alloy corresponds to a perfect flat interface between two domains of different species and on this type of substrate a trivial, one-dimensional, percolation takes place. However, for any finite temperature, the percolation probability of one-dimensional structures decreases for large samples becoming zero in the thermodynamic limit. In fact, when $L \rightarrow \infty$ the probability of having a defect on the otherwise straight domain wall preventing the occurrence of percolation goes to 1 if $T \neq 0$, which explains this result. On the other hand, when  the temperature is increased close to $T \simeq 2.80$ the density of dimers reaches a threshold that allows the onset
of percolating clusters. This fact will allow us to draw, in the phase diagram, a percolation line starting at $T \simeq 2.80$ that continues at higher temperatures.

Let us now analyze the properties of the percolation clusters. For this purpose we will first test the
size-scaling hypothesis for the fluctuations of the percolation threshold
and the extrapolation of that threshold with the system size, which are obtained in analogous
way that we have done
for $\theta_J$ and its fluctuations (see Equations
(\ref{thetax}) and (\ref{sigmax}), respectively).
The  fluctuations (RMS) of the $\theta_P$ given by $\sigma_{P}$
scale with the system size according to \cite{Stuffy}

\begin{equation}
\sigma_{P}\propto L^{-1/\nu_{P}}
\label{sigmaL}
\end{equation}

\noindent where the critical exponent $\nu_P$ is associated with
the divergent correlation length $\xi$ that behaves as

\begin{equation}
\xi\propto \vert \theta-\theta_{P} \vert ^{-\nu_P},
\label{longdiver}
\end{equation}

\noindent with $\nu_P = 4/3$ for random percolation \cite{Stuffy}.

\begin{figure}
\centerline{
\includegraphics[width=8cm,height=7cm,angle=0]{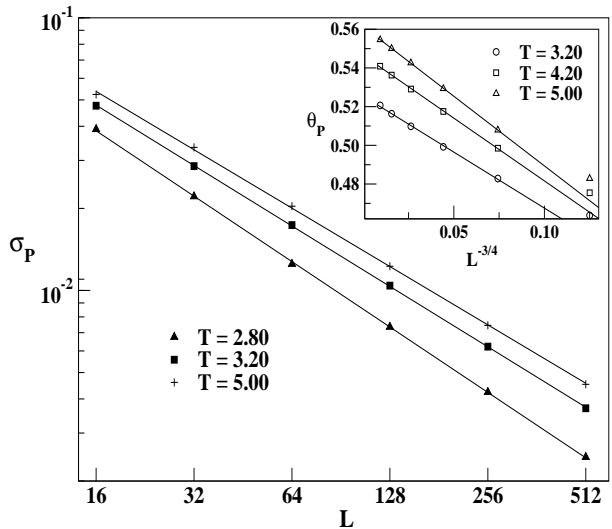}}
\caption{Log-log plots of $\sigma_{P}$ {\textit versus} $L$ at temperatures $T$ above the critical point. While the fitted exponents are consistent with standard percolation for most temperatures, there is a significant departure on the slope of the curve measured at $T$ = 2.80. This is the temperature at which the percolation line starts, with jamming and percolation occurring simultaneously in the thermodynamic limit.
The inset shows $\theta_{P}(L)$ {\textit versus} $L^{-1/\nu}$
at different temperatures $T$ above the critical point. The smallest lattice
($L=16$) has not been included in the fit.
}
\label{Fig5}
\end{figure}

Figure \ref{Fig5} shows \textit{log-log} plots of $\sigma_{P}$ \textit{versus}
$L$, obtained for fixed temperatures. The obtained values for
the exponent $\nu_P$ are compatible with standard percolation, as it follows from the data
listed in Table \textbf{I}. Only the exponent measured at
$T=2.80$ ($\nu_P^{-1} = 0.80 \pm 0.01$) falls below the expected value, suggesting that interesting physical processes may take place at the point where the jamming and percolation lines meet.
In order to extrapolate the percolation threshold to the
thermodynamic limit for $\theta_{p}(L)$ we can use the standard scaling
approach \cite{Stuffy} given by

\begin{equation}
\theta_{P}(L) = \theta_{P}({L\rightarrow\infty}) + B L^{-1/\nu_P} ,
\label{estrapolar5}
\end{equation}

\noindent where $B$ is a positive constant. The inset in Figure \ref{Fig5} shows plots of
$\theta_{P}(L)$ {\textit versus.} $L^{-1/\nu_{P}}$ obtained for $T > 2.80$ by taking
$\nu_P = 4/3$, as follows from the fit of the fluctuations
of $\theta_{P}(L)$ (see Equation (\ref{sigmaL}) and
Figure \ref{Fig5}). We found that Equation (\ref{estrapolar5}) holds and gives
a new confirmation of the value of the exponent $\nu_P$.
Furthermore, the fit allows us to extrapolate
the percolation threshold to the thermodynamic limit
for various temperatures (see also Table \textbf{I}).
At $T=2.80$ Equation (\ref{estrapolar5}) fits the data equally well for $\nu_P^{-1} = 3/4$
and $\nu_P^{-1} = 0.80$. Furthermore the extrapolated percolation threshold coincides, within error bars, for both exponents.

On the other hand, it is well known that, at the percolation threshold,
percolating clusters are objects with a well defined
fractal dimension $D_p$. The number of particles of the spanning
cluster in samples of side $L$ ($S(L)$) scales as

\begin{equation}
S(L) \propto L^{D_p}.
\label{dp}
\end{equation}

\begin{figure}
\centerline{
\includegraphics[width=7.5cm,height=6cm,angle=0]{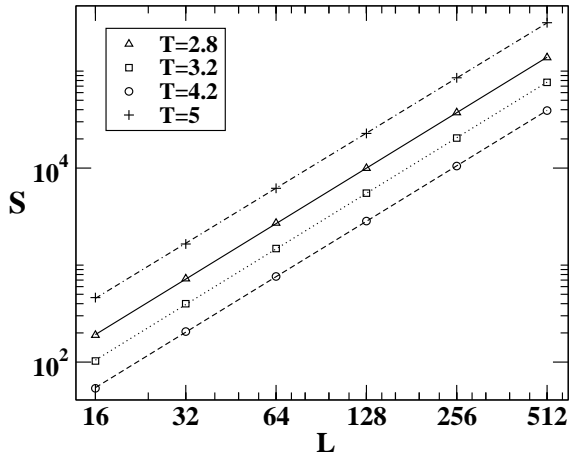}}
\caption{Log-log plots of $S(L)$ {\textit versus} $L$
at different temperatures $T$
above the critical point. We obtain the exponent $D_p\simeq1.89(1)$ corresponding to standard percolation in \textit{all} cases, including $T = 2.80$.}
\label{Fig6}
\end{figure}

Figure \ref{Fig6} shows log-log plots of the average
mass of percolating clusters {\textit versus} $L$, as obtained for
different annealing temperatures of the substrate.
In all cases (including $T=2.80$) the results obtained by fitting the data
with the aid of Equation (\ref{dp}) are in agreement with the fractal
dimension of standard percolation clusters given by
$D_p = 91/48 \approx 1.896$ \cite{Stuffy} (see Table \textbf{I}).
\\
\\
\textbf{Summary of the Results}
\\
\\
Figure \ref{Fig7} summarizes the results
obtained for jamming coverages and percolation thresholds in a phase diagram. Notice
that all values reported in Figure \ref{Fig7} correspond to
extrapolations to the thermodynamic limit performed with the aid
of Equations (\ref{estrapoteta}) and (\ref{estrapolar5}). The
following four regions can be observed in the phase diagram shown in Figure \ref{Fig7}:

\begin{figure}
\centerline{
\includegraphics[width=11cm,height=9cm,angle=0]{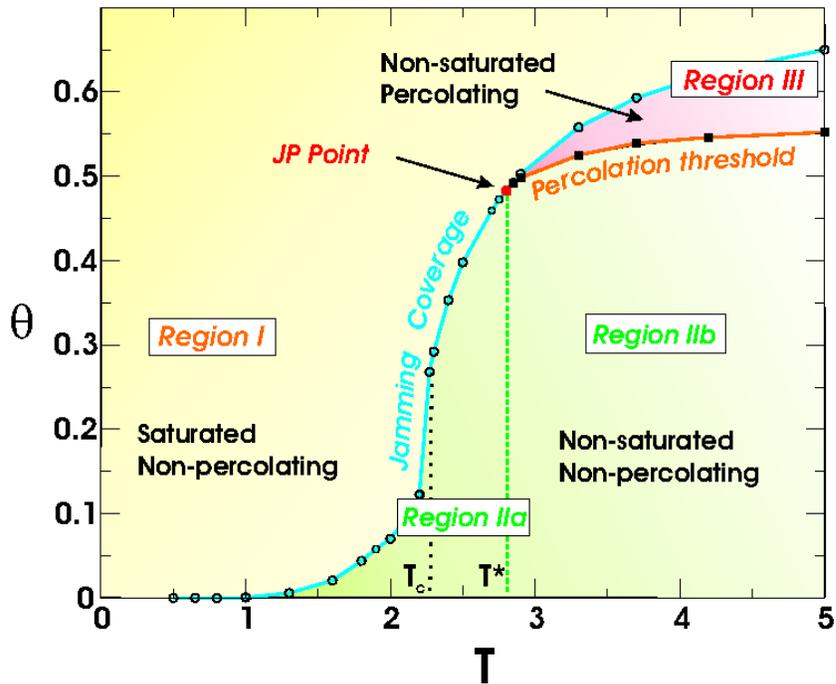}}
\caption{Phase diagram for percolation and jamming, summarizing the present results. We show the curves for jamming coverage $\theta_{J}$ (open circles) and the percolation threshold $\theta_{P}$ (filled squares) extrapolated to the thermodynamic limit {\textit versus} the annealing temperature $T$ of the alloy. The different regions are described in more detail in the text. In addition to the sharp changes taking place at the ordering temperature $T_c$ of the substrate, there exist peculiarities at the point where the jamming and percolation lines meet (point \textbf{JP}).}
\label{Fig7}
\end{figure}

\textbf{ Region I}: This corresponds to jammed states that
are inaccessible to the system. At low temperatures one has low
coverages since the sites of the substrate suitable for
dimer adsorption lie mainly along the interfaces between
domains of different particles. By increasing $T$
this scenario changes due to interdiffusion of species
causing the formation of additional $A-B$ pairs that implies an increase of the jammig coverages. This process
becomes particularly relevant close to the critical temperature
of the alloy ($T_{c} = 2.269$), so that for $T > T_{c}$ one has that
the jammed state is observed at higher coverages.

\textbf{ Region II}: Here the system has not saturated, but percolation is not observed. As shown in Figure \ref{Fig7}, we have divided {\bf Region II} into two parts:
\textbf{Region IIa}, limited by the jamming curve and a vertical-line at $T^* \simeq 2.80$; and \textbf{ Region IIb}, above $T^{*}$ but below the percolation line. The line of solid squares above $T^*$ marks the percolation threshold.

\textbf{Region III}: within this Region the coverages are high enough to observe
percolating clusters before the system gets jammed.

So far, along the percolation line the observed clusters belong to
the universality class of random percolation. It seems then that the universality
of the percolation process is not affected by the inhomogeneities
of the substrate annealed at different $T$. This finding is consistent
with the fact that the correlation length of the percolation process is
the only relevant length scale. However, inspection of
the phase diagram shows that there is a non-trivial point
at the intersection of all the regions (see the ``JP point'' in
Figure \ref{Fig7}). Furthermore, as we noticed before (Figure \ref{Fig5}) the behavior of the variance of the percolation threshold indicates a non-standard exponent.

\begin{table}
\begin{center}
\caption{
Critical percolation exponents measured at different temperatures
as listed in the first column. The number in parentheses represents an
estimation of the error in the last figure.
The 2nd and 3rd columns one show
the exponents obtained by fitting Equation (\ref{sigmaLT}) and Equation (\ref{dp}) to the simulation results, respectively.
The 4th column includes the extrapolated values of the obtained threshold percolation by mean of Equation  (\ref{estrapolar5}) using $1/\nu_p=3/4$.}
\begin{tabular} {|c|c|c|c|}
\hline
 Temperature & $1/\nu_p$  & $D_p$&   $\theta_p$   \\ \hline
 2.80        & $0.80(1)$   & $1.90(1)$   &   $0.483(1)$       \\ \hline
 3.20        & $0.74(1)$   & $1.89(1)$   &   $0.525(1)$       \\ \hline
 4.20        & $0.73(1)$   & $1.90(1)$   &   $0.546(1)$       \\ \hline
 5.00        & $0.74(1)$   & $1.89(1)$   &   $0.5613(5)$        \\ \hline
\end{tabular}
\end{center}
\end{table}

Since this finding may imply a change in the universality class for percolation at this particular point (in what seems to be an analogy with a multicritical point in thermodynamics), we have investigated it in particular in the next section, by introducing the concept of \textit {thermal percolation}.
\\
\\
{\bf IV. Thermal percolation.}
\\
\\
As it follows from the phase diagram shown in Figure \ref{Fig7}, it is possible to cross the \textbf{JP} point from a percolating region for $T > T^{*}$ to a non-percolating region
for $T < T^{*}$, just by moving along the jamming curve ($\theta_J(T)$)
sweeping the temperature. In this way the percolation probability ($P_L$) depends on  temperature, which in turn controls the properties of the substrates. In other words one has $P_L(\theta,T) = P_L(\theta_J(T),T) \equiv \Phi_L(T)$, in contrast to the standard approach where the density is the control parameter. We will now explore the validity of the scaling hypothesis in this new scheme.

\begin{figure}
\centerline{
\includegraphics[width=8	cm,height=6cm,angle=0]{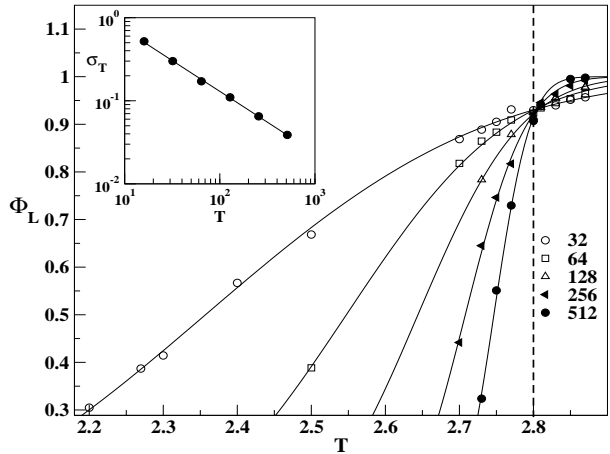}}
\caption{Plots of the percolation probability $\Phi_{L}(T)$ measured over the jamming curve
{\textit versus} $T$ as obtained using lattices of different size $L$
listed in the figure. $\Phi_{L}(T)$ is evaluated as the fraction of jammed deposition runs that have percolated for substrates annealed at temperature $T$. Solid lines correspond to fits of the numerical data to a Equation (\ref{errorT}). Note that all the curves intersect at $T = T^*$ and $\Phi^*=\Phi(T^*)\simeq0.93$, implying that jamming interferes with deposition for any value of $L$. The inset shows the width of the percolating transition when we take $T$ as the control parameter. The exponent we extract from the fit indicates a universality class that corresponds to standard percolation.}
\label{Fig8}
\end{figure}

Figure \ref{Fig8} shows typical curves of the $L-$dependent percolation
probabilities ($\Phi_{L}(T)$) {\textit versus} $T$ in a range of temperatures
close to $T^*$ = 2.80. It is found that curves corresponding
to different sizes have an unique intersection point given by
$\Phi^*=\Phi_L(T^*)\simeq0.93$.

\begin{figure}
\centerline{
\includegraphics[width=9cm,height=7cm,angle=0]{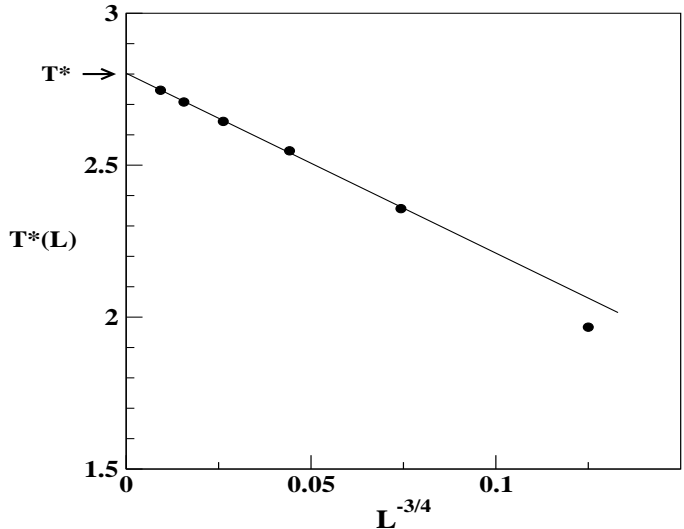}}
\caption{ Size-dependence of the critical temperature at
the percolation threshold $T^{*}_{L}$ {\textit versus} $L^{-1/\nu_{T}}$. In the fit (continuous line) we imposed $\nu_{T} = 4/3$.  The extrapolation to the thermodynamic limit gives $T^*(L\rightarrow\infty) = 2.80(1)$} (indicated in the graph by the arrows).
\label{Fig9}
\end{figure}
The shape of these curves, resembling so much those obtained for the percolating probability as a
function of the occupied fraction of sites, strongly suggests testing the finite-size scaling approach by using the temperature as a control parameter. In order to do this we first fit the curves of Figure \ref{Fig8} by means of an error function given by (see also Equation (\ref{error}))
\\
\begin{equation}
\Phi_{L}(T) = \frac{1}{\sqrt{2\pi}\sigma_{T}(L)} \int_{-\infty}^{T}
exp\Big{[}-\frac{1}{2}
{\Big{(}\frac{T^{'} - T^*(L)}{\sigma_{T}(L)}\Big{)}}^2
\Big{]}dT^{'}  .
\label{errorT}
\end{equation}
\\
In this way one obtains the thermal width of the transition
($\sigma_{T}$) and the critical threshold $T^*$. Of course,
both quantities depend on $L$. We now propose that thermal fluctuations should scale
with the size of the system in the same way as
density fluctuations do, namely following the analogous
to Equation (\ref{sigmaL}). So,
\\
\begin{equation}
\sigma_{T}\propto L^{-1/\nu_T} ,
\label{sigmaLT}
\end{equation}
\\
\noindent where $\nu_{T}$ is the correlation length exponent.
A log-log plot of $\sigma_{T}$ {\textit versus} $L$ (see
the inset in Figure \ref{Fig8}) shows that the scaling form of Equation (\ref{sigmaLT}) holds and by fitting the data we obtain $1/\nu_{T} = 0.74\pm0.01$ or, equivalently, $\nu_{T} = 1.35 \pm 0.02$. Quite remarkably, in contrast to the result obtained by studying percolation in the standard ensemble, the measured exponent for $T = T^*$ is in excellent agreement with the correlation length exponent of the standard percolation problem  ($\nu_p = 4/3$).

Now, the next step is to extrapolate the critical temperature---$T^{*}(L)$, which we have previously obtained using Equation (\ref{errorT})---to the thermodynamic limit by using
an Ansatz analogous to Equation (\ref{estrapolar5}), namely
\\
\begin{equation}
T^*(L) = T^*({L\rightarrow\infty}) +C  L^{-1/\nu_T} ,
\label{estrapolar2}
\end{equation}
\\
\noindent where $C$ is a constant. The obtained results are shown in Figure \ref{Fig9}. By using Equation (\ref{estrapolar2}), with $\nu_{T} = 4/3$ as it follows from the fit performed
to the data shown in inset of Figure \ref{Fig8}, we have determined a more accurate value of $T^*$ confirming that the critical temperature in the thermodynamic limit is given by $T^*(L\rightarrow\infty) = 2.80(1)$.

\begin{figure}
\centerline{
\includegraphics[width=10.5cm,height=8cm,angle=0]{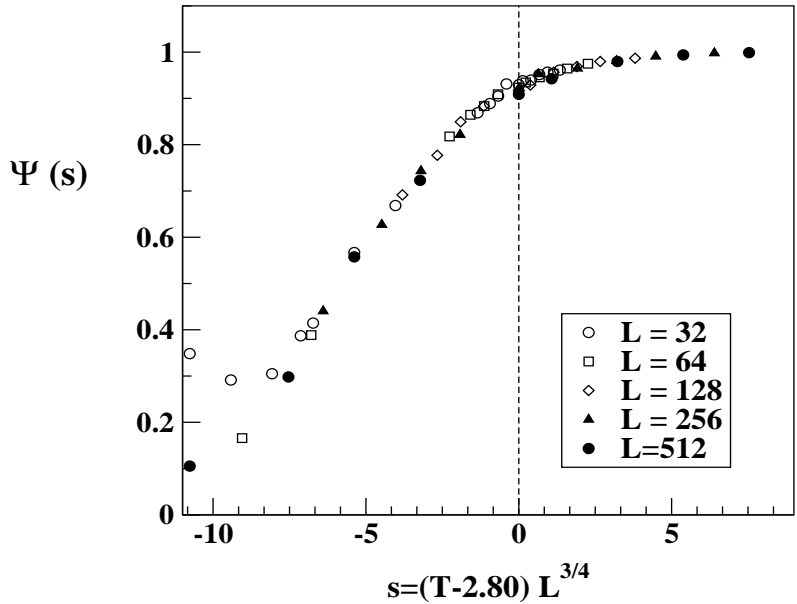}}
\caption{Scaled plots for the percolation probability $\Psi(s)$
{\textit versus} the reduced variable $s$ as obtained using lattices
of different size $L$. No free parameters were fitted to obtain the collapse. For additional details see the text.}
\label{Fig10}
\end{figure}

Finally, the scaling laws given by Equations (\ref{sigmaLT}) and
(\ref{estrapolar2}) in connection to Equation (\ref{errorT})
predict the collapsing of all the curves of $\Phi_{L}(T)$ shown in
Figure \ref{Fig8} when they are plotted as a function of a reduced
scaling variable $s \equiv (T-T^{*})L^{1/\nu_{T}}$.
In fact, Figure \ref{Fig10} shows a plot of the universal scaling function
$\Phi_{L}(T) = \Psi(s)$ that results from the collapse of data
corresponding to samples of several sizes
and obtained by using the already determined values of both $T^{*}$ and $\nu_{T}$ .
The quality of the collapse, obtained without any adjustable parameters,
is additional evidence of the validity of the proposed scaling Ansatz for thermal percolation.

In order to round out the present set of results two things remain to be explained: $i$) why we get a different exponent from the standard percolation value when the finite-size scaling behavior of the percolation coverage variance is analized close to $T$ = 2.80; $ii$) why we recover the usual exponent when we use $T$ as a tuning parameter for percolation moving along the jamming curve.

Regarding the first issue, we will show below that the interference between jamming and percolation forces a different exponent when measuring at fixed $T$ = 2.80. Indeed, at this temperature the threshold coverages for both phenomena are very close together for finite $L$, and they actually coincide for the infinite lattice. This implies that the fluctuations in the percolation coverage are restricted by the early onset of saturation. In other words, in certain stochastic deposition runs either jamming occurs too early or percolation too late, so that the system saturates before it percolates. Then, for a fixed $L$, we are measuring a \emph{reduced} value of $\sigma_P$, which depends on the distance $\mid\theta_J - \theta_P\mid$ relative to $\sigma_J$. The exponent for $\sigma_P$ that results from varying $L$ depends then on the interplay of these three quantities, giving a non-trivial---and erroneous---result. Of course, this interference is not operative when the percolation coverage is far apart from saturation ($\theta_J(L)-\theta_P(L) >> \sigma_{P}(L)$). So, in this case, we recover the normal exponents (see Table \textbf{I}).

In order to test these ideas we measured $\sigma_P$ at $T^*$ in another way, trying to remove the constraints imposed by jamming. We measured the probability of percolation $\Phi_{L}(\theta)$ at $T^*$ over 10-50 samples, and we fitted it with an error function (Equation (11)) normalized with the \textit{total} number of runs (~$10^5-10^6$). This is an important detail given that only a fraction of the runs has percolated (notice that the numerical data for $\Phi_{L}(\theta)$---Figure 9---never reach unity at any value of $\theta$ due to the fact that $\Phi_{L}(\theta)$ is fixed at $\Phi^*=0.93$ independent of the system size).
In this way, we obtained a corrected value of the variance for the percolation coverage $\sigma_P'(L)$. After fitting this corrected values with Equation (\ref{sigmaL}) we recovered the exponent $\nu$ that characterizes \emph{standard} percolation. We also tried constraining the fit of $\Phi_{L}(\theta)$ to values of $\theta$ well below $\theta_J$,  where the fraction of jammed depositions is negligible. Since, within error bars, we still got the same outcome as with the previous procedure (\emph{i.e.} $\nu \simeq 4/3$), we are quite confident of this result and the proposed explanation for the observed discrepancy.

Let us tackle now point $ii$): why do we measure $\nu$ = 4/3 when we move along the saturation curve through the percolation temperature threshold? In order to understand this, we will assume that when $T$ is varied in the range of temperatures studied in Figure  \ref{Fig8}, the connectivity changes in the underlying lattices and the adsorbed layer are not as important--concerning percolation--as the changes in coverage. On moving along the saturation curve (Figure 7) the coverage is fixed by the temperature $T$ (within a deviation $\sigma_J$ that is very narrow compared to $\sigma_P$ and gets narrower with increasing $L$ \cite{van}). We can then assume that $T$ and $\theta$ are almost interchangeable or essentially linked through a simple functional dependence. If now we accept that  $\theta_J$ is linear enough as a function of $T$ near $T^*$ we would then be measuring the probability of percolation in the usual way--at a given coverage $\theta_J(T)$--avoiding the interference between percolation and jamming. These arguments explain the scaling and the standard value for $\nu_{T}$ that we  obtained from the thermal analysis.
\\
\\
{\bf  V. Conclusions.}
\\
\\
Based on a numerical study of the random sequential adsorption of dimers on non-homogeneous binary alloys in the square lattice, we have shown that the jamming coverage (for temperatures below $T_c$) and its fluctuations (over the whole range of temperatures) show the same
size-scaling properties as percolation. However, the corresponding exponents are different, in spite of the fact that they depend on the same dimensions ($\nu = \frac{1}{D-d_f/\alpha}$ with $\alpha=1$ for the jamming coverage and $\alpha=2$ for its fluctuations). This diversity is in remarkable contrast to the case of percolation, where both exponents \textit{are} the same, being given by the divergence of the correlation length. We have also demonstrated that the incipient percolation cluster belongs to the universality class of standard percolation, as follows from the evaluated critical exponents through a finite-size scaling treatment of the numerical data. In this way inhomogeneities of the substrate are irrelevant for the
percolation phenomena.

In addition to these observations, an intersection between the jamming
and percolation curves is found. At this particular point we observed
a subtle interference between jamming and percolation, which seems to change one of the exponents associated with percolation. However, the constraint introduced by the jamming process on the percolation phenomena at this point not only does not change the universality class of the
percolation process but, as is shown, the jamming states at different $T$ can be used to characterize the critical behavior of  the percolating system. In this way, we have shown how to use the temperature as the control parameter that governs the percolation process.
We conclude that a generalization of the standard finite-size
scaling Ansatz formulated in terms of the density also holds true
when the control parameter is the temperature. In this way we are
able to characterize the percolation transition at the point at which the two lines intersect by showing that it still belongs to the standard
random-percolation universality class.


{\bf  ACKNOWLEDGMENTS}. This work was supported by UNLP,
CONICET and ANPCyT (Argentina).


%



\begin{thebibliography}{99}

\bibitem{evans} J. W. Evans, Rev. Mod. Phys. \textbf{65}, 1281 (1993).
\bibitem{reviews}P. Wero\'{n}ski, Advances in Colloid and Interface Science \textbf{118} 1 (2005) ; Z. Adamczyk, K. Jaszcz\'{o}lt, A. Michna, B. Siwek,
L. Szyk-Warszy\'{n}ska, M. Zembala Advances in Colloid and Interface Science \textbf{118}  25 (2005)
\bibitem{polaco} G. Kondrat, J. Chem. Phys. \textbf{124}  54713 (2006).
\bibitem{albanol} E. S. Loscar, and E. V. Albano, Rep. Prog. Phys. \textbf{66} 1343 (2003).
\bibitem{libros} \textit{The Chemical Physics of Solid Surfaces and Heterogeneous Catalysis, Vol. 4},
edited by D. A. King, and D. P. Woodruff (Elsevier, Amsterdam 1982).
\bibitem{priv} M. C. Bartelt, and V. Privman, J. Chem. Phys. \textbf{93}, 6820 (1990).
\bibitem{Don_04}  A. Donev, I. Cisse, D. Sachs, E. A. Variano, F. H. Stillinger, R. Connelly, S. Torquato, and P. M. Chaikin, Science \textbf{303} 990 (2004).
\bibitem{Mak_02} A. Donev, F. H. Stillinger, and S. Torquato, Phys. Rev. Lett. \textbf{95}  090604 (2005).
\bibitem{Stuffy} D. Stauffer, and A. Aharony,
\emph{Introduction to Percolation Theory} (Taylor and Francis, London, 1992).
\bibitem{Ber_04} L. Bergqvist, O. Eriksson, J. Kudrnovský, V. Drchal, P. Korzhavyi, and I. Turek, Phys. Rev. Lett. \textbf{93} 137202 (2004).
\bibitem{Mot_03} Y. Motome, N. Furukawa, and N. Nagaosa, Phys. Rev. Lett. \textbf{91} 167204 (2003).
\bibitem{perc2} G. Mackay, and N. Jan, J. Phys. A \textbf{17} L757 (1984)
\bibitem{Havlin} A. Bunde, and S. Havlin, \emph{Fractals and Disordered Systems} (Springer-Verlag. Berlin 1995).

\bibitem{Feders} J. Feders, \textit{Fractals}. (New York: Plenum Publishers, 1988).
\bibitem{van} N. Vandewalle, S. Galam, and M. Kramer, Eur. Phys. J. B, {\bf 14}, 407 (2000).
\bibitem{lba1} E. S. Loscar, R. A. Borzi, and E. V. Albano, Phys. Rev. E \textbf{68}, 041106 (2003).
\bibitem{lba2} E. S. Loscar, R. A. Borzi, and E. V. Albano, Eur. Phys. J. B \textbf{36}, 157 (2003).
\bibitem{lero} Y. Leroyer, and E. Pommiers, Phys. Rev. B {\bf 50}, 2795 (1994).
\bibitem{choi} H. S. Choi, J. Talbot, G. Tarjus, and P. Viot,
Phys. Rev. E {\bf 51}, 1353 (1995).

%
\bibitem{kon1} G. Kondrat, and A. Pekalski. Phys. Rev. E
{\bf 64}, 056118 (2001).
\bibitem{kon} G. Kondrat, and A. Pekalski. Phys. Rev. E
{\bf 63}, 051108 (2001).
\bibitem{fede} F. Rampf, and E. V, Albano. Phys. Rev. E
{\bf 66}, 061106 (2002).
\bibitem{naka} M. Nakamura, J. Phys. A
{\bf 19}, 2345 (1986).
\bibitem{evans2} J. W. Evans, and D. E. Sanders, Phys. Rev. B \textbf{39}, 1587 (1986).
\bibitem{polaco1} G. Kondrat, J. Chem. Phys. \textbf{122} 184718 (2005).
\bibitem{giorgio} M. Quintana, I. Kornhauser, R. L\'opez, A. J.
Ramirez-Pastor, and G. Zgrablich, Physica A {\bf 361} (2006) 195.
\bibitem{cecilia} M. C. Gimenez, F. Nieto, and J. Ramirez-Pastor, J. Phys. A: Math. Gen. {\bf 38}, 3253 (2005).
\bibitem{Binder} {\it The Monte Carlo Method in Condensed Matter Physics}, Edited by
K. Binder (Springer-Verlag, Berlin, 1992) .
\bibitem{stuffyb} D. Stauffer, Physica A {\bf 242} (1997) 1.
%
\end{thebibliography}
\end{document}